\documentclass[conference]{IEEEtran}
\IEEEoverridecommandlockouts

\usepackage{cite}
\usepackage{amsmath,amssymb,amsfonts}
\usepackage{algorithmic}
\usepackage{graphicx}
\usepackage{textcomp}
\usepackage{xcolor}
\def\BibTeX{{\rm B\kern-.05em{\sc i\kern-.025em b}\kern-.08em
    T\kern-.1667em\lower.7ex\hbox{E}\kern-.125emX}}
\begin{document}

\title{Integrating Deep Learning for Arrhythmia Detection with Automated Drug Delivery: A Comprehensive Approach to Cardiac Health Monitoring and Treatment
}

\author{\IEEEauthorblockN{Vatsal Sivaratri}
\IEEEauthorblockA{\textit{Thomas Jefferson High School for Science and Technology} \\
Alexandria, USA \\
vatsal.sivaratri@gmail.com}
\and
\IEEEauthorblockN{Praveen Kumar Pandian Shanmuganathan}
\IEEEauthorblockA{\textit{Florida Institute of Technology} \\
Melbourne, Florida \\
ppandianshan2015@my.fit.edu}
}

\maketitle

\begin{abstract}
Arrhythmias are irregularities in the heart’s electrical system which cause rapid and irregular heartbeats \cite{b1}. These heart conditions affect over 33 million people globally and significantly increase the risk of severe complications, including stroke, heart failure, and sudden death \cite{b2}. Modern screening and treatment approaches, like 12-lead ECG tests and analyzing patient medical history, use frameworks that don’t address early onset of conditions and lack sufficient information to optimize treatment plans post-diagnosis. This project aimed to enhance cardiac arrhythmias' early diagnosis, monitoring, analysis, and treatment using an optimized 5-step patient pathway. We developed deep learning models using ECG, PPG, and SpO2 data to monitor conditions remotely with smartwatches and document arrhythmic episodes with relevant information, including daily patterns. We synthesized these into patient reports suitable for real-world clinical, providing enough information to guide treatment decisions before any new diagnostics. For critical care in a hospitalized setting or personalized home care, we developed a novel drug delivery system synchronized with a phone via Bluetooth and uses prescription-based or prediction-based to deliver medication intravenously at the optimal time. This approach could reduce the need for additional diagnostic tests, streamline patient management, and optimize medication schedules to better align with the individual's physiological needs, significantly improving patient outcomes.
\end{abstract}


\section{Background}
Arrhythmias are characterized by an abnormal rhythm of the heart, which can lead to various health challenges and risks. The prevalence of arrhythmias varies, with conditions like atrial fibrillation affecting up to 2\% of the global population- a figure that increases with age \cite{b3}. Arrhythmias can arise from multiple causes, such as ischemic heart disease, hypertension, valve disorders, and congenital heart defects, which disrupt the normal conduction pathway of the heart's electrical impulses \cite{b4}. In individuals over 65 years old, the prevalence can be as high as 10\%, making it a major contributor to morbidity in aging populations. Other types of arrhythmias, such as ventricular fibrillation (VFib), can be fatal within minutes if not treated, highlighting the need for timely intervention.

The impact of arrhythmias can range from symptoms like palpitations, dizziness, and fainting, to more serious complications such as stroke (particularly in atrial fibrillation) and sudden cardiac arrest (as in ventricular fibrillation). Given the serious risks, treatment options such as medication, lifestyle changes, and interventions like pacemakers or defibrillators are essential in managing arrhythmias and reducing associated health risks.

Electrocardiography (ECG) is a fundamental diagnostic tool for identifying arrhythmias, as it captures the heart's electrical activity through electrodes placed on the skin. These electrodes detect the subtle electrical signals generated by the depolarization of cardiac muscle cells during each heartbeat, allowing clinicians to analyze heart rhythm and identify abnormalities. Different types of arrhythmias produce characteristic ECG patterns, making the technique indispensable for diagnosis \cite{b5}. This non-invasive test can reveal a wide range of conditions, from benign arrhythmias to life-threatening disturbances, such as atrial fibrillation and ventricular tachycardia.

In addition to ECG, photoplethysmography (PPG) is gaining prominence as a non-invasive method for monitoring heart rhythm. PPG uses a light source and a photodetector positioned on the skin to measure blood volume changes. The light is absorbed differently depending on the blood pulse, and these fluctuations are captured by the photodetector to calculate heart rate and rhythm. PPG is particularly advantageous for continuous, passive monitoring and has proven useful in wearable technologies that can detect early arrhythmic episodes without the need for clinical supervision \cite{b6}. Its integration into consumer health devices offers a practical method for early detection, although PPG is generally seen as a complement to, rather than a replacement for, ECG due to limitations in accuracy under certain conditions.

The treatment of arrhythmias varies according to the type and severity of the condition. Pharmacological approaches often involve antiarrhythmic drugs, which work by altering the electrical conduction in the heart to restore or maintain normal rhythm. In more severe cases, electrical cardioversion or catheter ablation may be necessary. Electrical cardioversion delivers a controlled electric shock to reset the heart's rhythm, while catheter ablation involves targeting and destroying the specific heart tissue that is causing the abnormal electrical signals. In cases of recurrent or life-threatening arrhythmias, the implantation of pacemakers or defibrillators can be life-saving. Pacemakers regulate slow heart rhythms, while implantable cardioverter-defibrillators (ICDs) deliver shocks to correct dangerous fast rhythms like ventricular fibrillation \cite{b7}, \cite{b8}.

This project aims to improve existing healthcare pathways for patients with heart arrhythmias by building on current diagnostic and treatment methodologies. It leverages the accuracy of gold-standard techniques such as ECG and clinician verification while addressing key bottlenecks in the patient journey. These include enhancing the accuracy of diagnostics with current hardware, providing more detailed information about the patient's condition before the first appointment, and optimizing drug delivery based on individual needs. This novel approach could significantly improve early detection, treatment personalization, and overall patient outcomes, paving the way for more efficient arrhythmia management.

\section{Methods and Materials}
\subsection{Dataset}
Data for this project was collected from three primary sources, with the key focus on ECG (electrocardiogram) and PPG (photoplethysmogram) recordings of patients diagnosed with various types of arrhythmias. Two main datasets were identified that fit these criteria: the MIMIC PERForm AF (Atrial Fibrillation) Dataset \cite{b9}, available on Zenodo, and A Large-Scale 12-Lead Electrocardiogram Database for Arrhythmia Study hosted on PhysioNet \cite{b10}.

The MIMIC PERForm dataset consists of 20-minute ECG and PPG recordings from critically ill adult patients during routine clinical care. Of the recordings, 19 were captured during episodes of atrial fibrillation (AFib), while 16 represented normal sinus rhythm, all sampled at 125 Hz. This dataset is a curated subset of the broader MIMIC III Waveform Database, which is widely known for its extensive collection of physiological signals from ICU patients.

The second dataset, A Large-Scale 12-Lead ECG Database, offers a collection of 45,152 ECG recordings, covering more than 55 different arrhythmia classifications. Each recording is 10 seconds long and sampled at a higher frequency of 500 Hz. For this project, only a subset of the data was used. Certain arrhythmia classes were excluded, and the remaining were grouped into four broader categories: Normal Sinus Rhythm, Atrial Fibrillation, Bradycardia, and Tachycardia.

\begin{figure}[htbp]
\centerline{\includegraphics[width=0.5\textwidth]{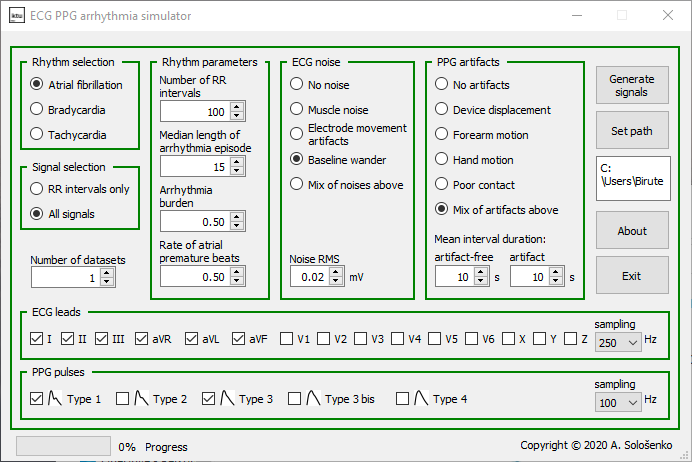}}
\caption{Arrhythmia Episode Simulator}
\label{fig}
\end{figure}

These datasets provided a rich foundation for analyzing arrhythmic patterns and developing models for arrhythmia detection and classification. The inclusion of both high-frequency, real-time data from critically ill patients and extensive multiclass arrhythmia labels allowed for a comprehensive exploration of the electrical anomalies in the heart.

In addition to leveraging existing datasets, we incorporated an Arrhythmia episode simulator to generate supplementary PPG data (Figure 1) \cite{b11}, \cite{b12}, \cite{b13}, \cite{b14}. 

This simulator is capable of replicating a range of arrhythmias, including Atrial Fibrillation (AFib), Bradycardia, and Tachycardia. It allows for flexible control over key parameters, such as episode duration, RR interval count, and the number of simulated recordings. Furthermore, the simulator accounts for the physiological imperfections commonly found in real-world waveform data, offering the ability to introduce noise, and artifacts, and adjust the frequency of premature beats. While testing these features, we observed that higher levels of artificially generated noise did not align with the noise patterns seen in real clinical data. Excessive noise caused the simulated recordings to diverge too distinctly from those of normal sinus rhythm, compromising the realism of the dataset. To maintain a balance between authenticity and variability, we opted for a low artifact-to-clean data ratio of 30:1. Based on this configuration, we generated 50 episodes each of AFib, Bradycardia, and Tachycardia, with 100 RR intervals per episode and a median episode length of 15 intervals.

This carefully calibrated approach ensured that the simulated data remained representative of real-world physiological conditions while enhancing the robustness of our model by accounting for the variability seen in clinical recordings.

\subsection{Patient Pathway Design}

\begin{figure}[htbp]
\centerline{\includegraphics[width=0.5\textwidth]{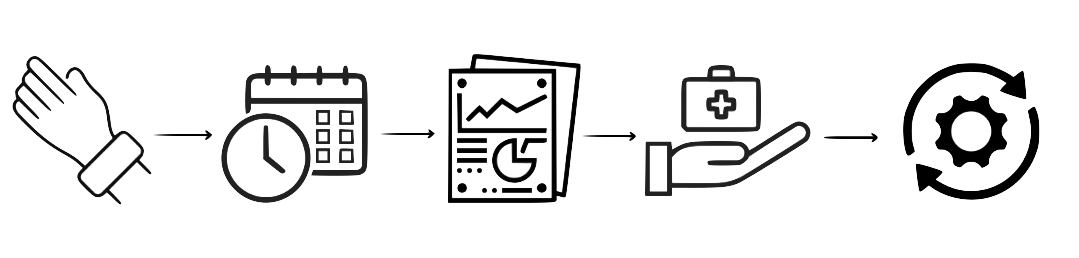}}
\caption{Arrhythmia Episode Simulator}
\label{fig}
\end{figure}

To effectively address the limitations of existing workflows, it is crucial to demonstrate that the proposed solutions can be integrated into practical patient care pathways. In this paper, we present a five-step pathway designed to address a wide range of arrhythmias. While the specific details of treatment will vary based on the patient's condition, these steps represent the core structure that can be applied in most cases.
 
The first step is screening the patient for the risk of arrhythmia. Current treatment methods for arrhythmias often rely on patients experiencing symptoms, which is suboptimal and, in some cases, life-threatening. To mitigate this, we propose using existing smartwatch technology—such as the Apple Watch, Samsung Galaxy Watch, and FitBit—to continuously monitor for arrhythmia risk. While Apple has already incorporated a system for detecting irregularities, particularly Atrial Fibrillation, other major manufacturers have yet to implement official, comprehensive systems. Moreover, there is room for improvement in classification granularity, allowing for the detection of a broader range of arrhythmias.

In the first phase of testing, a PPG sensor will continuously monitor the patient for irregular heart activity. The PPG sensor measures changes in blood volume in the wrist, which correlates with the heart's electrical activity and blood circulation. The data used for these models comes from the MIMIC PERForm AFib dataset and a data simulator. This approach provides an initial assessment of heart irregularities.

The second step involves using a 1-lead ECG to more precisely identify the type of arrhythmia a patient may have. For this task, we categorize the data into four main groups: Healthy Control, Atrial Fibrillation, Bradycardia, and Tachycardia. We use the large-scale 12-Lead ECG dataset, focusing on Lead I to simulate data that can be collected by a smartwatch. In practice, the patient will be continuously monitored via PPG, with the smartwatch prompting them to take ECG readings during times of heightened arrhythmic episodes. This combination of PPG and ECG data provides more precise insights into the patient's condition.

Once the system determines whether the patient should consult a healthcare professional, the next critical step is collecting additional data for further evaluation. After scheduling an appointment, the patient will be asked to take ECG readings throughout the day, which will be logged as part of a report. This continuous monitoring helps doctors assess the development of the arrhythmia before the actual appointment. Additionally, a questionnaire will gather clinically relevant information, including demographics, past medical history, and other important metrics. Key metrics include the HAS-BLED score, which assesses the risk of major bleeding, and the CHA2DS2-VASc score, which evaluates stroke risk in patients with Atrial Fibrillation. These detailed reports enable doctors to better understand the patient's condition and minimize the risk of misdiagnosis or improper treatment selection.

After the initial appointment and any additional tests, doctors can prescribe medication based on the patient’s condition. However, two significant challenges remain: medication adherence and treatment optimization. Medication adherence is a major concern, with studies showing that only around 50\% of patients with chronic illnesses follow their prescribed treatments \cite{b15}. Additionally, many treatments are not tailored to individual needs, resulting in higher long-term costs and poorer outcomes.

The final and most critical step in this pathway is the implementation of timed drug delivery. Using ECG and PPG sensors, the proposed system will identify the optimal time to administer medication based on the patient's arrhythmia cycles. Evidence suggests that arrhythmias often follow circadian rhythms, allowing for the prediction of episodes and the timing of medication to preempt these events. The system will also ensure adherence to the doctor's prescribed treatment plan, preventing issues such as toxicity (e.g., administering 5 mL of medication three times a day). This approach optimizes treatment effectiveness and reduces the likelihood of complications due to poor adherence or improper medication timing.

This five-step approach provides a structured, technology-enhanced pathway that improves arrhythmia management, from early detection and diagnosis to treatment and follow-up care.

\subsection{Spectrogram Creation and Model Selection}

Spectrograms are graphical representations that show how the frequency content of a signal changes over time. They are typically generated by breaking down a signal into smaller segments and analyzing the frequency components within each segment. This time-frequency analysis provides valuable insight into the dynamic behavior of signals, especially when the signal contains complex patterns, non-stationary elements, or time-varying frequency characteristics.

To create a spectrogram, methods like the Continuous Wavelet Transform (CWT) or Short-Time Fourier Transform (STFT) are often employed. Each method has its own advantages. The CWT is particularly useful for capturing localized changes in a signal and works well for signals where both high and low-frequency components need to be captured at different resolutions. It provides a multi-resolution analysis, offering finer time resolution for high-frequency components and better frequency resolution for low-frequency components. We used the CWT technique to create a spectrogram for our experiment. 

\begin{figure}[htbp]
\centerline{\includegraphics[width=0.4\textwidth]{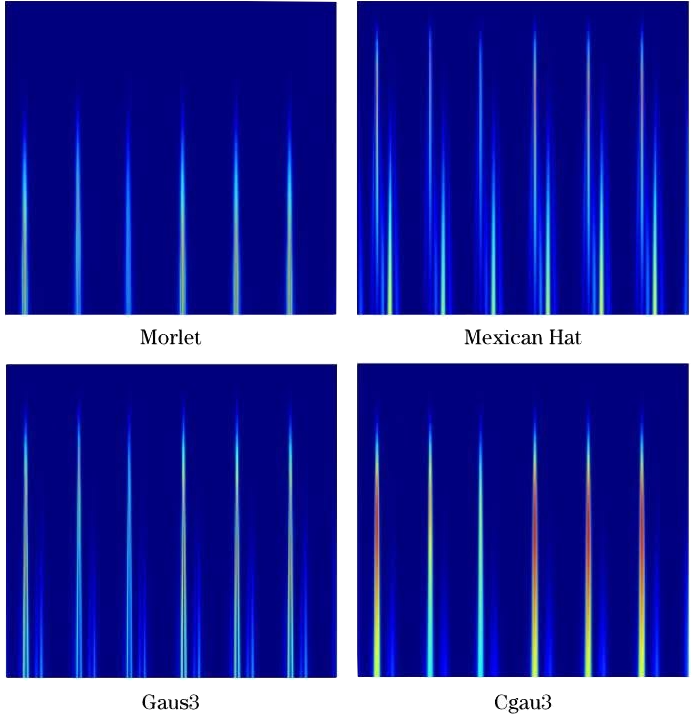}}
\caption{Arrhythmia Episode Simulator}
\label{fig}
\end{figure}

The Morlet wavelet \cite{b16}, employed in our study for ECG signal analysis, is a complex wave that combines a cosine function with a Gaussian window. This unique formulation makes it highly effective for analyzing non-stationary signals like ECGs, where frequency content can change over time.

\begin{equation}
\Psi(t) = e^{i 2 \pi f_c t} e^{-\frac{t^2}{2 \sigma^2}}
\end{equation}

In this equation:
\begin{itemize}
    \item \( e^{i 2 \pi f_c t} \) represents the sinusoidal (cosine) wave, where \( f_c \) is the central frequency and \( t \) is time. This component provides the frequency information.
    
    \item \( e^{-\frac{t^2}{2 \sigma^2}} \) is the Gaussian window, where \( \sigma \) controls the width of the window and ensures the wavelet is localized in time. This window tapers the sinusoidal wave, limiting the wavelet to a specific time range.
\end{itemize}
The modulation by the Gaussian window allows the wavelet to be localized in time, which is crucial for analyzing non-stationary signals like ECGs. The Gaussian window smooths the edges of the sinusoidal wave, ensuring that the wavelet captures frequency information for specific time intervals rather than over an infinite duration.

This combination of time and frequency localization makes the Morlet wavelet a powerful tool for time-frequency analysis, as it enables the detection of transient changes in the frequency content of the signal, which is important for identifying patterns and anomalies in ECG data.

It is beneficial for arrhythmia detection tasks, because of its high resolution in both time and frequency domains. In the context of ECG signals, it can accurately determine the location and frequency of transient events, such as R-peaks, which are important for determining different types of arrhythmia. The clear spectral lines of the Morlet wavelet, as illustrated in the attached image, provide distinct patterns that improve the accuracy of our models.

We used CNNs due to their proven effectiveness in image classification tasks, which can be leveraged for time-series classification of ECG signals treated as one-dimensional "images". By processing the data through multiple convolutional layers, CNNs can learn to identify features that are indicative of different arrhythmic conditions.

\subsection{Prototype and Experiments}
For our hardware drug delivery system, we developed, 3D-printed, and assembled a functional prototype (Figure 4). The design prioritizes both functionality and user comfort, with the 3D-printed components ergonomically shaped to fit around the patient’s arm. This wearable device houses several critical components, including a Raspberry Pi 5, a stepper motor, and a custom-designed rack and pinion mechanism, which collectively work to automate the movement of the syringe for precise drug administration.

\begin{figure}[htbp]
\centerline{\includegraphics[width=0.4\textwidth]{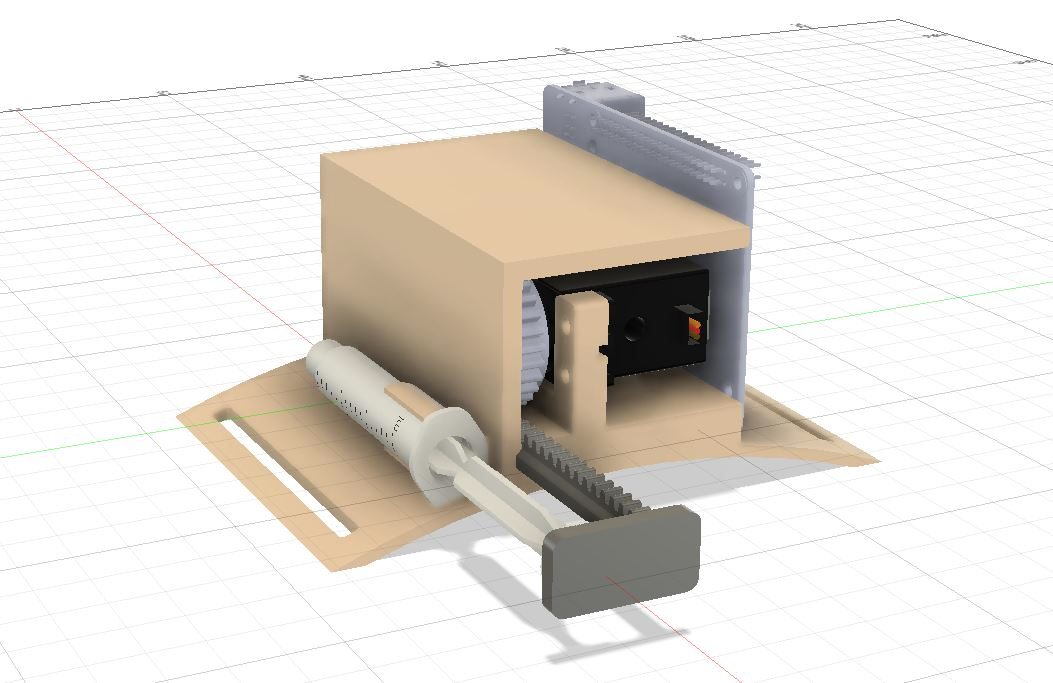}}
\caption{3D CAD Model Prototype Drug Delivery System}
\label{fig}
\end{figure}
The Raspberry Pi 5 serves as the central controller for the system and is configured to respond to incoming requests through a web server interface. To facilitate remote control of the drug delivery, we developed a Flask-based application, which hosts a web server on a designated port. This allows the system to be easily accessible via client devices such as smartphones or smartwatches, enabling users to trigger drug delivery on demand. The application communicates with the hardware to execute precise movements of the syringe via the stepper motor, ensuring that the correct dosage is administered at the appropriate time.

The custom rack and pinion arrangement plays a crucial role in ensuring smooth, accurate, and consistent movement of the syringe plunger. This mechanical setup converts the rotational motion of the stepper motor into linear motion, allowing the device to control the syringe’s displacement with a high degree of precision. The stepper motor’s incremental movements ensure that the exact dosage of medication is delivered without error, making it suitable for applications where accurate drug administration is critical, such as in arrhythmia management or timed medication delivery.
\begin{figure}[htbp]
\centerline{\includegraphics[width=0.3\textwidth]{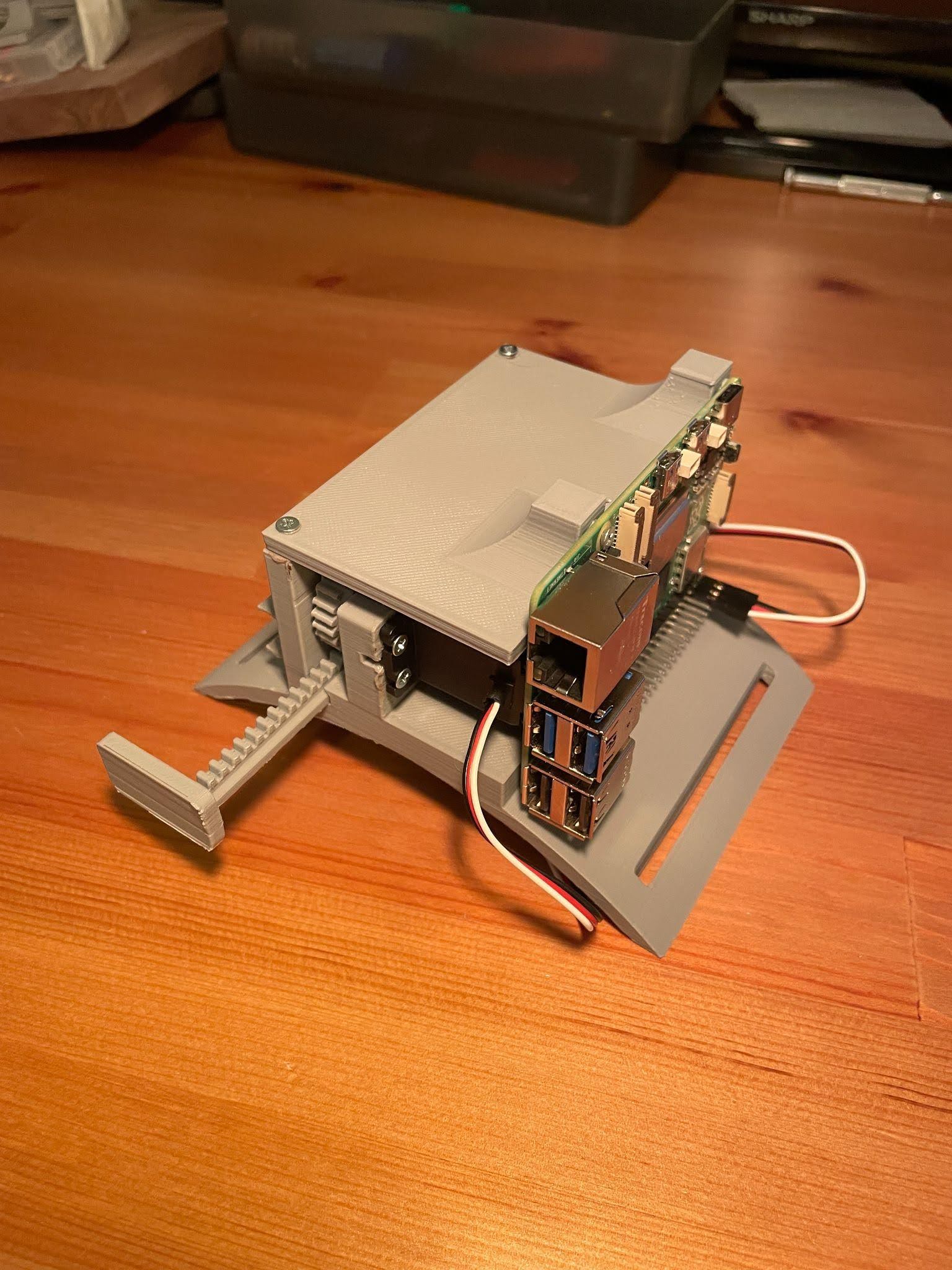}}
\caption{3D Printed Prototype Drug Delivery System}
\label{fig}
\end{figure}
The integration of a web server allows for remote control and monitoring of the drug delivery system. By using a smartphone or smartwatch, the patient or healthcare provider can initiate drug delivery as needed, or the system can be programmed to automatically administer medication at predetermined times, aligning with the patient's arrhythmic episodes or circadian rhythms. This flexibility ensures that medication can be tailored to the patient’s individual needs, providing a personalized treatment experience.

Additionally, the Raspberry Pi 5 and Flask application offers the potential for future improvements and customization. In future improvements, we plan to integrate additional monitoring features, adjust the timing of drug delivery, or implement safety protocols based on specific medical requirements using potentially conducting a clinical trial. The combination of accessible hardware and programmable software makes the system adaptable to a wide range of medical scenarios, ensuring that the drug delivery is both reliable and responsive to the patient’s condition.

In addition to the hardware prototype, we also made Neural Network models to detect arrhythmia events using ECG and PPG data inputs. The results from the models are provided in detail in the next section. 
\section{Results}
The results from Figure 6 demonstrate that the DenseNet121 model achieved the highest accuracy among the ECG models tested, with an impressive precision of 0.996 and an Area Under the Curve (AUC) of 0.997, indicating excellent overall performance. However, although MobileNetV2 had the fastest average processing time, it exhibited lower accuracy and AUC compared to DenseNet121. This suggests that despite MobileNet's similar accuracy in some cases, not all transfer learning models are well-suited for this specific task. Therefore, it is essential to carefully evaluate which models are most effective for ECG analysis.

\begin{table}[htbp]
\caption{ECG Model Results}
\begin{center}
\renewcommand{\arraystretch}{1.5}
\resizebox{\columnwidth}{!}{
\begin{tabular}{|l|c|c|c|c|c|c|c|c|}
\hline
\textbf{Model} & \textbf{Avg Time (s)} & \textbf{Accuracy} & \textbf{Specificity} & \textbf{Precision} & \textbf{Recall} & \textbf{F1 Score} & \textbf{PRC} & \textbf{AUC} \\
\hline
DenseNet121 & 0.305439 & 0.985 & 0.995444 & 0.99568 & 0.997836 & 0.996757 & 0.992131 & 0.996894 \\
EfficientNetB0 & 0.185772 & 0.975 & 0.993182 & 0.99352 & 0.995671 & 0.994595 & 0.987955 & 0.994984 \\
MobileNet & 0.137963 & 0.976667 & 0.997712 & 0.997826 & 0.993506 & 0.995662 & 0.983855 & 0.993393 \\
MobileNetV2 & 0.131437 & 0.861667 & 0.880282 & 0.900585 & 1 & 0.947692 & 0.855453 & 0.934678 \\
ResNet50 & 0.254369 & 0.975 & 0.986364 & 0.987179 & 1 & 0.993548 & 0.976744 & 0.990644 \\
Xception & 0.251206 & 0.983333 & 0.995465 & 0.99568 & 0.997836 & 0.996757 & 0.985038 & 0.993734 \\
\hline
\end{tabular}
}
\end{center}
\label{tab1}
\end{table}

In Figure 7, the PPG models show an even higher level of accuracy, with both DenseNet121 and Xception models reaching a perfect AUC score of 1.000. This suggests that the PPG models may be particularly adept at distinguishing between different types of arrhythmic events, and are highly “confident” in their decisions. The consistency across accuracy, precision, recall, and F1 score across all models further reinforces the potential of PPG data in arrhythmia detection, which then sets up the context for classification.

\begin{table}[htbp]
\caption{PPG Model Results}
\begin{center}
\renewcommand{\arraystretch}{1.5}
\resizebox{\columnwidth}{!}{
\begin{tabular}{|l|c|c|c|c|c|c|c|c|}
\hline
\textbf{Model} & \textbf{Avg Time (s)} & \textbf{Accuracy} & \textbf{Specificity} & \textbf{Precision} & \textbf{Recall} & \textbf{F1 Score} & \textbf{PRC} & \textbf{AUC} \\
\hline
DenseNet121 & 0.259945 & 0.998095 & 0.995962 & 0.996407 & 1 & 0.9982 & 0.998788 & 0.999317 \\
EfficientNetB0 & 0.173405 & 0.998095 & 0.995962 & 0.996407 & 1 & 0.9982 & 0.998792 & 0.999323 \\
MobileNet & 0.125461 & 0.998095 & 0.995962 & 0.996407 & 1 & 0.9982 & 0.998788 & 0.999321 \\
MobileNetV2 & 0.127021 & 0.993016 & 0.993271 & 0.993983 & 0.992788 & 0.993385 & 0.999247 & 0.998719 \\
ResNet50 & 0.241151 & 0.998095 & 0.997308 & 0.997599 & 0.998798 & 0.998198 & 0.999958 & 0.999953 \\
Xception & 0.251058 & 0.99873 & 0.997308 & 0.997602 & 1 & 0.9988 & 1 & 1 \\
\hline
\end{tabular}
}
\end{center}
\label{tab1}
\end{table}

\begin{figure}[htbp]
\centerline{\includegraphics[width=0.3\textwidth]{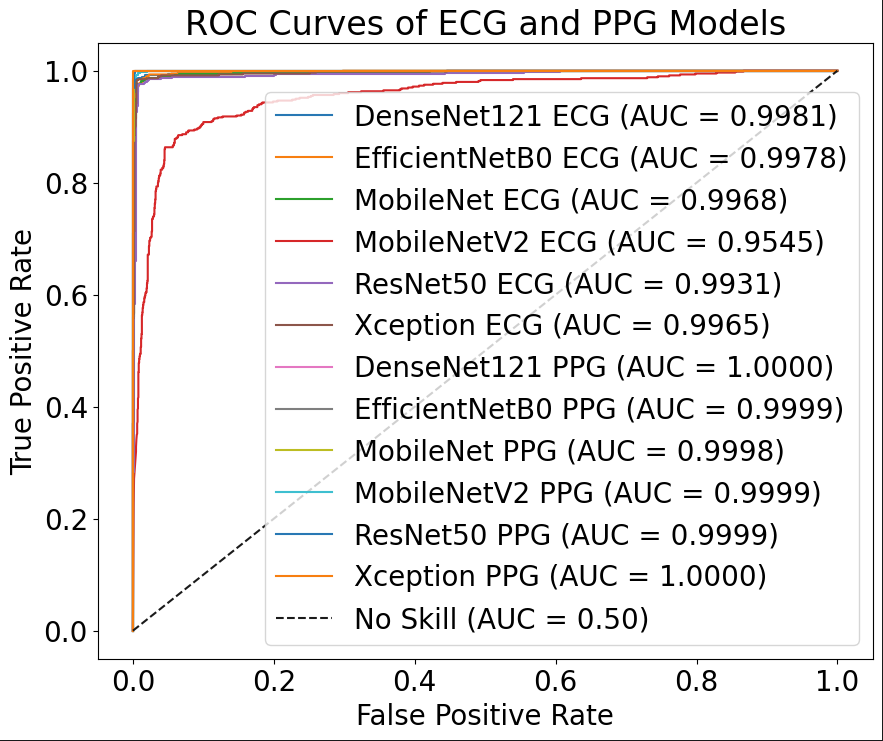}}
\caption{PPG Model Results}
\label{fig}
\end{figure}

Figure 8 presents the Receiver Operating Characteristic (ROC) curves for both ECG and PPG models, with the curves of the PPG models, particularly DenseNet121 and Exception, aligning closer to the top-left corner. This alignment signifies a good true positive rate without a significant increase in the false positive rate. The AUC values near or equal to 1.000 for PPG models reinforce the reliability of these models in correctly classifying arrhythmic events. It should be noted that the Area under the curve computed in the graph is not consistent with that of the corresponding ECG table, and this is due to Scikitlearn’s implementation varying from that of Tensor Flows. At any rate, both measurements show the same relative differentiation of model performance, meaning one can regard either of them as valid.

\begin{figure}[htbp]
\centerline{\includegraphics[width=0.3\textwidth]{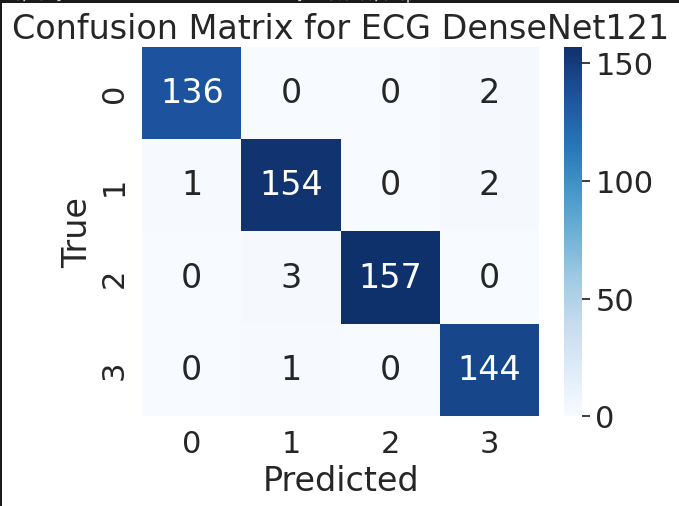}}
\caption{ECG DenseNet121 Confusion Matrix}
\label{fig}
\end{figure}

The Confusion Matrix in Figure 9 for the ECG DenseNet121 model illustrates the model's high sensitivity and specificity, with only a single false negative and very few false positives. 

\section{Conclusion}

This study demonstrates the effectiveness of deep learning models in accurately classifying arrhythmias using both ECG and PPG data, confirming their potential for real-world application in cardiac health monitoring. The models achieved high accuracies and consistently reliable results, which underscores their capability to detect and classify arrhythmia. The ability of these models to learn from large datasets and generalize across different types of arrythmic events opens the door for integration into wearable health devices, which can enable continuous, non-invasive monitoring for patients at risk of heart conditions.

By leveraging deep learning techniques, these models can analyze complex patterns in ECG and PPG signals that may not be easily detectable through traditional methods. This enables early detection and real-time monitoring of arrhythmias, which is crucial for timely medical intervention. For example, continuous monitoring through wearable devices like smartwatches, equipped with these algorithms, could alert patients and healthcare providers to irregular heart rhythms long before symptoms appear, thereby reducing the risk of complications such as stroke or heart failure.

In addition to the performance of the deep learning models, this study also validated the functionality of the proposed hardware system for drug delivery, further advancing the potential for personalized treatment in cardiac care. The 3D-printed prototype, designed to fit ergonomically around the patient’s arm, successfully integrated a Raspberry Pi 5 to manage the operation of a stepper motor and a custom rack-and-pinion mechanism for precise syringe movement. One of the standout features of this system is its low response time, which is crucial for real-time medical applications. This rapid responsiveness ensures that medication can be delivered promptly during critical moments, such as when arrhythmic episodes are detected.

The platform-agnostic nature of the web-based control means that it can be integrated into existing healthcare infrastructures with minimal modifications, allowing healthcare providers to remotely monitor and control the drug delivery process, potentially even automating it based on the patient's real-time cardiac data. The versatility of the hardware system extends beyond its use in arrhythmia management. The ability to administer drugs via a syringe mechanism controlled by a stepper motor makes it suitable for intravenous (IV) drug delivery, where precise control over medication dosage and timing is critical. This feature could be especially valuable in hospital settings or for patients requiring home-based care, as it allows for continuous, accurate medication delivery without requiring constant manual intervention by healthcare providers. This can potentially be expanded to treat other illnesses and conditions like Asthma, Bronchiectasis, Cystic Fibrosis, etc.

In conclusion, this study not only confirms the potential of deep learning models for arrhythmia classification using ECG and PPG data but also introduces a novel hardware solution for personalized and automated drug delivery. The combination of accurate arrhythmia detection with a responsive and flexible drug delivery system paves the way for more integrated, patient-centered approaches to managing cardiac conditions. As both the algorithms and hardware are refined, this technology has the potential to revolutionize how we monitor, diagnose, and treat heart diseases in both clinical and home-based environments.

\end{document}